Opinion

# Real-Time Dynamic Optimal Power Flow in Electric Vehicles Considering the Lifetime of the Components in the E-Powertrain

**Erfan Mohagheghi[1]*, Joan Gubianes Gasso[1] and Pu Li[2]**

[1]MicroFuzzy GmbH, Taunusstraße 38, 80807 Munich, Germany

[2]Department of Process Optimization, Ilmenau University of Technology, Ilmenau, Germany

***Corresponding author:** Dr. Erfan Mohagheghi, MicroFuzzy GmbH, Taunusstraße 38, 80807 Munich, Germany, E-mail: Erfan.Mohagheghi@microfuzzy.com

Different types of energy sources (e.g., batteries, supercapacitors, fuel cells) can be utilized in electric vehicles to store and provide energy in the e-powertrain through power electronic devices [1-6]. The lifetime of the components in the e-powertrain depends on their load profile [7,8]. For instance, the lifetime of a battery highly depends on the depth of discharge and the number of charge/discharge cycles [9-13]. The lifetime of an inverter mostly depends on the variations in the active-reactive power passing through it. This means that the expended life cost of the components can be decreased by allocating an optimal share of the total power to each energy source and power electronic device at an optimal time instance. In addition, the driving range of a vehicle can be prolonged by decreasing the energy loss i.e., operating the components in their high-efficiency region. Therefore, it is necessary to perform Optimal Power Flow (OPF) in the operation of the e-powertrain. The OPF aims to minimize the expended life cost of the components and maximize the driving range of the vehicle by optimizing the following decision variables while satisfying technical constraints:

- The charge/discharge power of battery storage and supercapacitor systems 'at each time instance'
- The charge power of fuel cell systems at each time instance
- The load share of each DC/DC converter
- The bidirectional active and reactive power profiles of each DC/AC inverter
- The length of charge and discharge periods of batteries and supercapacitors
- The number of charge-discharge cycles of each storage unit in each prediction horizon
- The status of charge/discharge of batteries and supercapacitors
- The depth of discharge for storage systems at each time instance

This leads to a large-scale (with hundreds of variables), non-convex, stochastic (due to uncertain parameters), dynamic (due to storage components), multi-timescale, and mixed-integer nonlinear programming (MINLP) problem. Thus, the computation time to solve the problem can be much higher than required for 'real-time' application [14,15]. In addition, the feasibility of the real-time solutions should be also ensured for the safe operation of the vehicle [16]. For this reason, MicroFuzzy GmbH, in collaboration with the Technische Universität Ilmenau, develops a multi-time-horizon framework to solve this challenging optimization problem in real-time (Figure 1). The computation time is decreased using parallel computing to achieve the online OPF. The solutions also safeguard both feasibility and optimality of the operations in real-time, while minimizing the total operation costs.

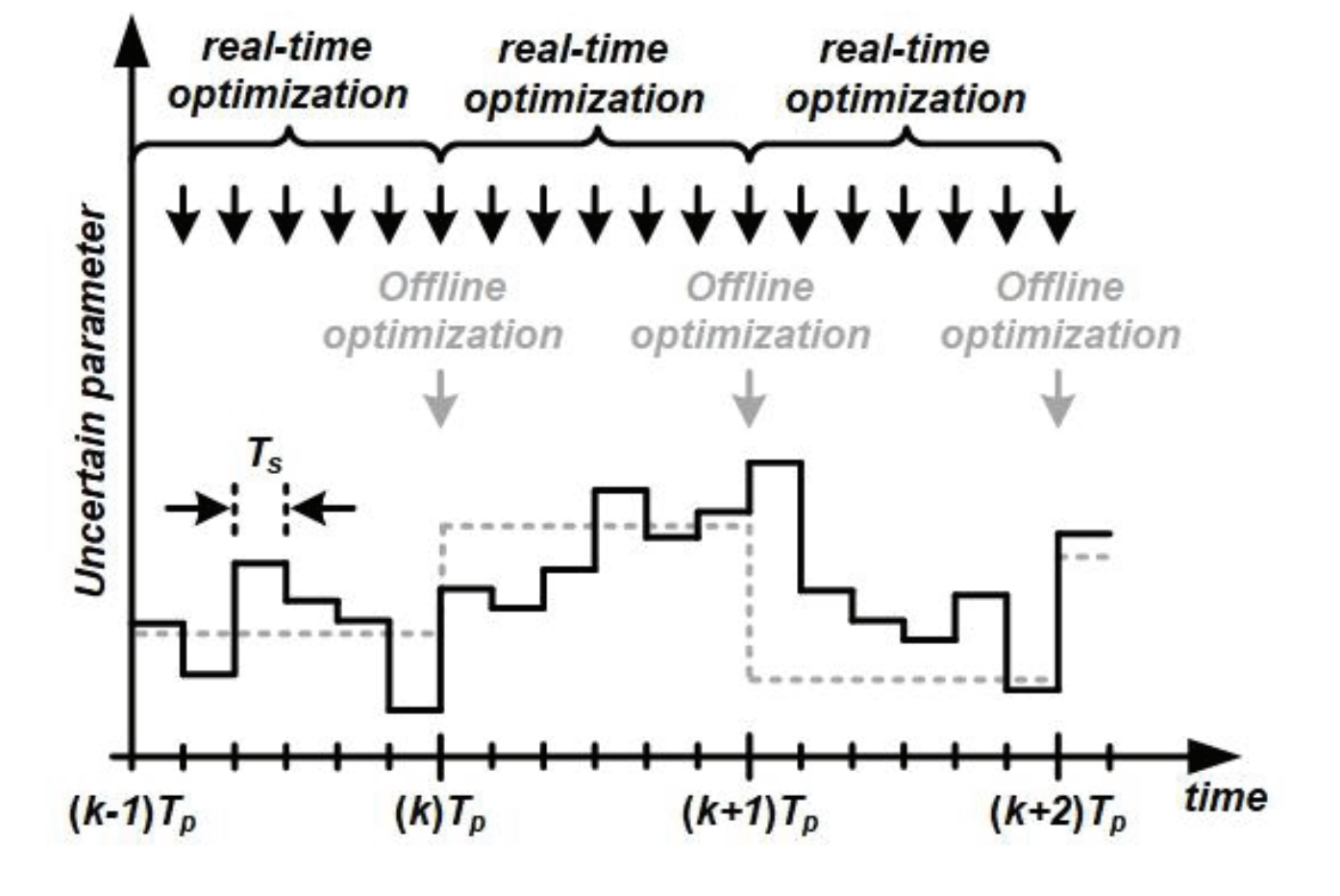

**Figure 1:** Offline optimal power flow (gray) and online optimal power flow (black).